\documentclass[12pt,titlepage]{utarticle}

\newcommand{\tr}{{\rm tr}}
\DeclareMathSymbol{\square}{\mathord}{AMSa}{"03}
\DeclareMathSymbol{\rightsquigarrow}{\mathrel}{AMSa}{"20}

\newdimen\tableauside\tableauside=1.0ex
\newdimen\tableaurule\tableaurule=0.4pt
\newdimen\tableaustep
\def\phantomhrule#1{\hbox{\vbox to0pt{\hrule height\tableaurule width#1\vss}}}
\def\phantomvrule#1{\vbox{\hbox to0pt{\vrule width\tableaurule height#1\hss}}}
\def\ZZ{\relax{\sf Z\kern-.4em Z}}
\def\sqr{\vbox{%
  \phantomhrule\tableaustep
  \hbox{\phantomvrule\tableaustep\kern\tableaustep\phantomvrule\tableaustep}%
  \hbox{\vbox{\phantomhrule\tableauside}\kern-\tableaurule}}}
\def\squares#1{\hbox{\count0=#1\noindent\loop\sqr
  \advance\count0 by-1 \ifnum\count0>0\repeat}}
\def\tableau#1{\vcenter{\offinterlineskip
  \tableaustep=\tableauside\advance\tableaustep by-\tableaurule
  \kern\normallineskip\hbox
    {\kern\normallineskip\vbox
      {\gettableau#1 0 }%
     \kern\normallineskip\kern\tableaurule}%
  \kern\normallineskip\kern\tableaurule}}
\def\gettableau#1 {\ifnum#1=0\let\next=\null\else
  \squares{#1}\let\next=\gettableau\fi\next}

\input epsf
\newcount\figno
\def\fig#1#2#3{
\par\begingroup\parindent=0pt\leftskip=1cm\rightskip=1cm\parindent=0pt
\baselineskip=11pt
\global\advance\figno by 1
\epsfxsize=#3
\centerline{\epsfbox{#2}}
\vskip 12pt
{\bf Figure \the\figno:} #1\par
\endgroup\par
}
\def\figlabel#1{\xdef#1{\the\figno}}
\def\encadremath#1{\vbox{\hrule\hbox{\vrule\kern8pt\vbox{\kern8pt
\hbox{$\displaystyle #1$}\kern8pt}
\kern8pt\vrule}\hrule}}

\begin{document}

\preprint{
 HUB-EP-97/27\\
 {\tt hep-th/9705022}\\
}

\title{Branes and Six Dimensional Fixed Points}
\author{Ilka Brunner and Andreas Karch
 \oneaddress{
  \\
  Humboldt-Universit\"at zu Berlin\\
  Institut f\"ur Physik\\
  Invalidenstra{\ss}e 110\\
  D-10115 Berlin, Germany\\
  {~}\\
  \email{brunner@qft1.physik.hu-berlin.de\\
  karch@qft1.physik.hu-berlin.de}
 }
}
\date{May 5, 1997}

\Abstract{We analyze brane configurations corresponding
to non-trivial six dimensional fixed points. Several
results previously obtained from a pure field theoretical
analysis are rederived in the brane language.}

\maketitle
\section{Introduction}
Recently, the world volume of branes was used to study field theories
in various dimensions with different amount of supersymmetry
\cite{hw,ahha,EKG,EKG2,brodie,barbon,shapere,theisen}.
The original construction in \cite{hw} was invented to study field theories in
2+1 dimensions with $N=4$ supersymmetry. The main idea used in that
paper  is to stretch $N_c$ parallel 3 branes of IIB string
theory between NS 5 branes and to study the world volume theory
of the 3 branes, which is an $SU(N_c)$ gauge theory. Matter is included
by introducing additional D5 branes, which give rise to fundamental
fields when they intersect the 3 branes.
This setup was modified to
study 4 dimensional $N=1$ theories in \cite{EKG}. In that
paper Seiberg's duality was interpreted in the brane setup.
The introduction of orientifolds into the setup of \cite{EKG}
enabled the authors of \cite{shapere} to study also $SO$ and $Sp$
groups.
In \cite{brodie}
many more 4d, $N=1$ field theory phenomena such as product gauge groups
and superpotentials were given an interpretation in a brane picture.
$d=4$ theories with $N=2$ are studied in \cite{witten}. Here also
quantum effects are interpreted in the brane setup by considering
the bending of the higher dimensional branes when the lower dimensional
branes end on them. In \cite{ahha} the construction of 5-dimensional
$N=1$ theories via branes is discussed. The authors construct
new 5-dimensional
superconformal field theories. Their paper contains also results on
$d=3$, $N=2$ theories, which were also studied in \cite{EKG2}.
In this paper, we will study 6-dimensional $N=1$ field theories
in the brane setup. In section 2 several aspects of the brane constructions
in various dimensions are reviewed. The 5-dimensional theories
of \cite{ahha} are discussed in some more detail in section 3. Here,
we also check the consistency of the construction in \cite{ahha}
using orientifolds.
Section 4 contains the 6-dimensional analysis. We show how
6-dimensional $SU, \,SO$ and $Sp$ gauge theories with a tensor
multiplet can be described in the brane picture. We compare our
results to the field theory results of \cite{sweden} and \cite{seiberg}.

\section{Hanany-Witten in various dimensions}

The Hanany-Witten \cite{hw} setup yields an $N=4$ supersymmetric theory in three
dimensions. Its basic ingredients are IIB NS 5 branes, D5 branes and
D3 branes. The worldvolumes of the various branes occupy the
following directions:

\begin{center}
\vspace{.6cm}
\begin{tabular}{|c||c|c|c|c|c|c|c|c|c|c|}
\hline
&$x^0$&$x^1$&$x^2$&$x^3$&$x^4$&$x^5$&$x^6$&$x^7$&$x^8$&$x^9$\\
\hline
NS 5&x&x&x&x&x&x&o&o&o&o\\
\hline
D 5&x&x&x&o&o&o&o&x&x&x\\
\hline
D 3&x&x&x&o&o&o&x&o&o&o\\
\hline
\end{tabular}
\vspace{.6cm}
\end{center}

In the $x^6$ direction the 3 branes are suspended between 5 branes. The
effective low energy theory is the 2+1 dimensional field theory
living on the extended 3 brane directions. Fluctuations of the
larger branes are considered much heavier than those of the 3 branes.
Moving around the higher dimensional branes hence corresponds to
changing parameters of the low energy theory, moving around the
3 branes corresponds to changing moduli of the theory. The
idea behind this approach can be summarized as the statement, that
the low energy dynamic is determined by the lowest dimensional brane in
the setup.

The simplest realization of the Hanany-Witten setup is a configuration
of $N_c$ 3 branes suspended between two NS 5 branes. An analysis
of the boundary conditions at the intersection of the branes shows
that this configuration describes a pure $N=4$ supersymmetric $SU(N_c)$
gauge theory in three dimensions. There are two ways
to incorporate flavors into this theory. One is to add D5 branes.
Whenever a D5 brane approaches the stretched D3 branes we obtain
a massless flavor hypermultiplet from strings stretching between
the D5 and the D3 brane. The other approach is to add semi-infinite
3 branes stretching to the right of the right or to the left of the left
NS brane \cite{witten}.
The flavor multiplets arise from 3-3 strings stretching
from a semi-infinite 3 brane to the finite 3 branes.
Since the worldvolume of these semi-infinite branes has
4 infinite directions they do not participate in the low energy dynamics
according to the general philosophy that only the lowest dimensional 
objects determine the low energy effective theory. Their motion again
corresponds to parameters of the theory. 
Latter approach has the advantage that it is in general easier to
deal with those configurations since we avoid introducing the third
kind of brane. But we have to pay a price. Using
semi-infinite 3 branes makes the Higgs branch of the theory very hard
to see. \footnote{We can change a flavor from a D5 brane to
a flavor from a semi-infinite 3 brane by moving the D5 brane off to
infinity. Higgs branches are associated to motion of D3 branes
connecting the flavor giving D5 branes with the various other branes.
Moving off the flavor giving D5 brane to infinity also moves the
Higgs branches off to infinity as well.}

There are two ways to generalize this setup. One is to rotate
the 5 branes \cite{EKG,brodie,barbon}. This reduces the supersymmetry from $N=4$ to $N=2$. The other
is to apply T-duality to a direction in the NS 5 brane worldvolume
\footnote{A NS 5 brane only stays a NS 5 brane under T-dualities in
one of its worldvolume directions}
and transverse to the D3 brane \cite{witten}. The resulting configuration
is a IIA D4 brane stretched between the two NS branes. Flavors
can again be added by incorporating D6 branes or semi-infinite
D4 branes. The low energy theory is a four dimensional
$N=2$ theory. A combination of these two approaches
has been pioneered in \cite{EKG} to yield a $d=4$, $N=1$ theory.

It is natural to T dualize once more to get $d=5, N=1$ from
IIB D5 branes still stretched between the two NS branes. Note
that this is not possible in the rotated (hence $N=1$) theory
since we would T dualize in a direction transverse to the rotated NS brane.
This is consistent with the fact that $N=1$ is the minimal supersymmetry 
in 5 dimensions. This step has been taken by \cite{ahha}. We will
review the parts of their construction relevant
for our analysis in the next section.

One can T dualize once more to obtain IIA 6 branes suspended
between the two NS 5 branes. This will be the subject of this
paper.

In all the configurations obtained by T-dualizing the original
Hanany-Witten setup, there is an unbroken $SO(3)$ subgroup of
the 10d Lorentz group corresponding to rotations in the 789 plane.
This corresponds to the $SU(2)_R$ symmetry present in all theories
with the equivalent of $N=2$ in $d=4$.

\subsection{Bending of branes}

From what we have said so far it is clear that
the configurations we are dealing with are ``color'' D d branes suspended
between two NS 5 branes of IIA(B) theory for even(odd) d, giving
rise to a d-1 dimensional low energy field theory. Even
though we always considered the 5 branes as being flat, this
classical picture is not the full story, since the d brane ending
on the 5 brane creates a dimple. The form of this deformation
depends crucially on d.

The end of the d brane looks like a $d-1$ brane in the
NS brane worldvolume or as a charged particle in
the $6-d$ dimensional subspace of the NS worldvolume transverse
to the d brane worldvolume. Generalizing the description
of this phenomenon in \cite{witten} one would like to describe
the bending of the NS brane as an equation
$$x^6 = x^6 ( {\bf y})$$
where ${\bf y}$ denotes a vector in the 6-d dimensional 
{\it transverse NS space}. Since
in ${\bf y}$ space the end of the d brane looks like a charged particle
we expect $x^6$ to satisfy the following equation:
$$ \Delta x^6 = 0$$
away from the ends of the d branes.
The behaviour in various dimensions can be analyzed by looking
at the Green's function of the Laplacian in the corresponding
$6-d$ dimensional transverse NS space.

The net charge is proportional to the number of d branes ending
on the left of a given NS brane minus the number of d branes ending
on the right, since they contribute charges of opposite sign.

\begin{itemize}
\item $d=3$: the Green's function in the 3 dimensional transverse
NS space goes like $x^6=\frac{1}{|{\bf y}|}+constant$. Far
away from the disturbance the position of the NS brane approaches
a definite value. It makes sense to talk about the position
of the NS brane. The inverse coupling constant is given
by the distance between these brane positions.

\item $d=4$: the Green's function in 2 dimensions yields
a logarithm. This is the case discussed in \cite{witten}. 
The distance between the two NS branes does not reach
a finite value but diverges logarithmically for large $|{\bf y}|$.
Witten interprets this as the sign of asymptotic freedom in the
brane picture. The distance in {\bf y} sets
the energy scales one is probing. Asymptotic freedom
is manifest through the logarithmic divergence of the
inverse gauge coupling. The coefficient
in front of the logarithm is given by the difference of
the net charges of the two NS branes. This
coincides with the one-loop beta function of
the corresponding field theory.

\item $d=5$: in 1 dimension we have a linear Green's
function yielding piecewise straight branes. The
gauge coupling is piecewise linear as expected from the
field theory analysis. We will say more about this case in the next section.

\item $d=6$: we can not have a net charge in 0 dimensions. The
brane describes a consistent theory only if the net charge on every
5 brane vanishes.
\end{itemize}

\section{The 5d analysis}
\subsection{The building blocks}

Let us briefly review the analysis of \cite{ahha} of the 5 dimensional
setup. We are dealing with NS 5 branes still along the 012345
coordinates and D5 branes long 012346, stretched between the
NS branes in the 6 direction. (This convention
differs slightly from the one used in \cite{ahha}, where they take the
NS brane to live in 012389 and the D5 in 012789. Our notation
is the one which should be obvious if one constructs this setup by
applying T-duality to the original Hanany-Witten setup.)
Flavors will be incorporated by semi-infinite 5 branes. In 5d the
alternative would be including D7 branes in the 01234789 direction.
The introduction of such objects would lead to new complications,
since these objects affect the asymptotic space-time geometry. So we
avoid them by using the semi-infinite 5 branes to add the flavors.
As a trade back, we loose the brane realization of the Higgs branches,
as mentioned above.

Next let's discuss the bending of the branes in more detail. We already
saw in the last section, that we are always dealing with straight
branes. They change their orientation, when they hit another 5 brane.
So one has to deal with branes living in the full 01234 space and describing
straight lines in the 56 plane. An NS brane only lives in the 5 and a D brane
only in the 6 direction. We can classify them by their charge under
the 2-form gauge potentials present in the IIB setup. The NS
brane is usually called a (1,0) brane and the D5 brane is the (0,1) brane.
There are also bound states of D and NS branes, (p,q) branes for
arbitrary p and q relatively prime. They behave exactly like ordinary 5 branes, since they are related by a $SL(2,\ZZ)$ duality transformation. (np,nq) branes
can be viewed as n copies of a (p,q) brane lying on top of each other.
Such a (p,q) brane also lives in the full 01234 space. The line
it describes in the 56 plane is given by the equation
$$ p x^6 + q x^5 = constant.$$
Charge conservation tells us that whenever n (0,1) branes end on the left
of m (1,0) branes, the m (1,0) branes will be "bended" to a (m,n) brane
(for m,n not relatively prime, this again has to be interpreted as
several branes lying on top of each other). This agrees with
the solution of the Laplace equation with a source term of strength
$\frac{n}{m}$.
An example of this effect is shown
in figure 1, which we borrowed from \cite{ahha}.

\vspace{1cm}
\fig{A D5 brane which ends on a NS 5 brane. The left side
describes the naive configuration, and the right side the correct
configuration, which implements conservation of charge at the vertex.
Figure borrowed from \protect \cite{ahha}.}
{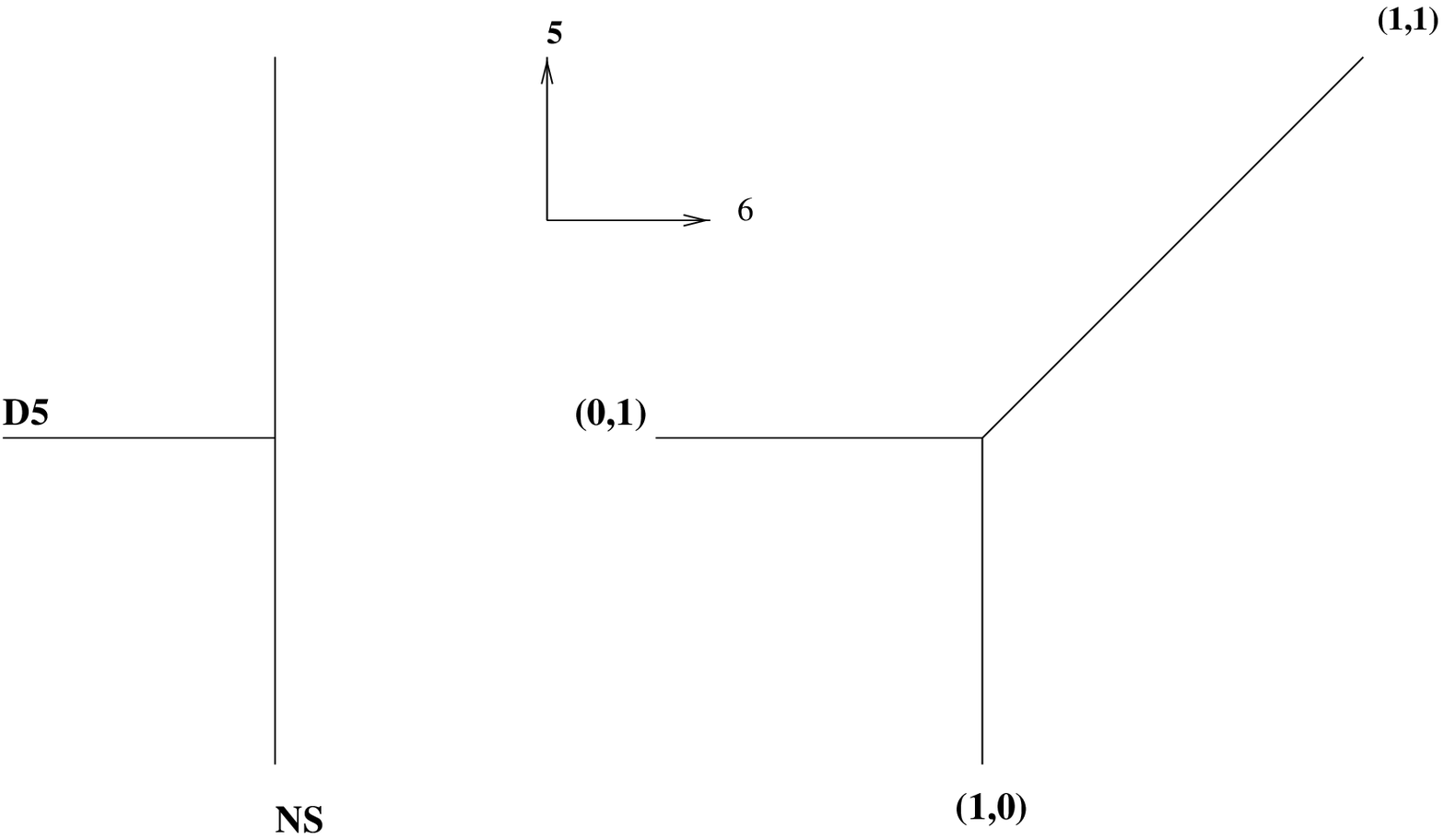}{15 truecm}
\figlabel\NSD
\vspace{1cm}

\subsection{The low energy field theory}

According to the general philosophy behind the field theoretic analysis
one should now look for the lowest dimensional objects in this setup,
since they determine the low energy field theory description.
In all the lower dimensional examples these objects were the
color d branes suspended between the NS branes. Here the situation
becomes more complicated: all finite 5 brane pieces should appear
on an equal footing. Only the semi-infinite 5 branes are heavy and
do not participate in the low energy dynamics. Hence
in this setup brane motions that leave the asymptotic form of the
5 branes invariant should be viewed as moduli of the theory while
a change in the asymptotic behaviour of the semi-infinite 5 branes
corresponds to changing the parameters of the theory.

To analyze this in more detail, let us start as in \cite{ahha} with
the setup that naively would correspond to a pure $SU(2)$ gauge theory:
Two D5 branes suspended between two NS branes. Taking
into account the linear bending, the situation is summarized in figure 2,
which we borrowed from \cite{ahha}. Even though
we expect that we get contributions from the finite
(1,0) pieces as well as from the finite (0,1) pieces, it is
easy to see that there is only one deformation of this
configuration which does not change the asymptotic forms
of the branes, corresponding to shrinking the rectangle
homogeneously in the 5 and 6 direction and thus going from configuration
(a) to configuration (b). This one parameter deformation
can still be interpreted as the Coulomb branch of an $SU(2)$ gauge theory.

\vspace{1cm}
\fig{Pure $SU(2)$ gauge theory in five dimensions. Horizontal lines represent
D5 branes, vertical lines represent NS 5 branes, and diagonal lines
at an angle $\theta$ such that $\tan(\theta)=p/q$ represent $(p,q)$
5 branes. Figure (a) shows a generic point on the Coulomb branch,
figure (b) shows a point near the origin of moduli space, and figure (c)
corresponds to the strong coupling fixed point. Figure borrowed from \protect
\cite{ahha}} {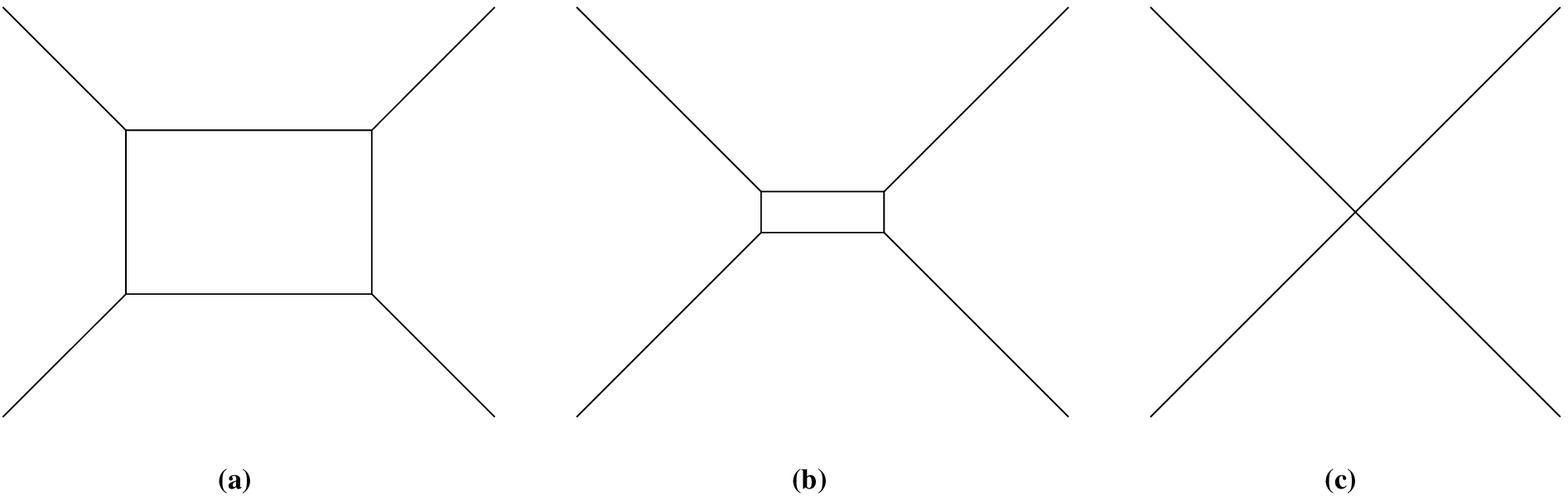}{13 truecm}
\figlabel\SUtwozero
\vspace{1cm}

When the two D5 branes meet, that is the rectangle degenerates to
a line, we expect the full $SU(2)$ symmetry to be restored.
The distance between the two vertices, where the semi-infinite
5 branes meet corresponds to the inverse bare gauge coupling $1/g_0^2$.
For finite $g_0^2$ this corresponds to a trivial IR fixed point.
For infinite $g_0^2$ this theory was argued to have a non-trivial
fixed point in \cite{sei5d}. This fixed point corresponds to figure (c). This 
situation can be generalized to the claim of \cite{ahha} that
every configuration of n semi-infinite half 5 branes emanating from
the same point correspond to a five dimensional IR fixed point. Some
of these fixed points may be trivial (for example for n$=$2). In the
rest of this paper we will assume that this is not the
case for n$>$2. It
is also an open question, if all fixed points generated this way are actually
distinct.

\subsection{The fixed point theories}

Aharony and Hanany studied the possible deformations of these fixed
points. Of special interest are deformations that correspond to Coulomb
branches coming out of the fixed point, like in the $SU(2)$ example
reviewed above. If after going to finite coupling these
Coulomb branches include only areas bounded
in the 5 direction by $N_c$ parallel 
D5 branes, these branches have a natural interpretation as Coulomb
branches of an $SU(N_c)$ gauge theory.
In addition there are several fixed points which have
Coulomb branches which can not be associated to Coulomb branches of
an gauge theory in any obvious way and fixed points which do
not have any Coulomb branches at all. There are no deformations
leaving the heavy branes untouched
corresponding to the Higgs branches of the gauge theories. This
is not too surprising, since we produced our flavors by
semi-infinite branes. This moved the Higgs branches off to infinity,
as discussed earlier. They are associated to motion of semi-infinite branes
which appear as parameters of the theory, not as moduli.

Given this brane realization of non-trivial IR fixed points,
an interesting question to ask is: given a field theory with
certain gauge group and matter content, does it
allow a non-trivial fixed point when we take the coupling to
infinity. This question was analyzed by Intriligator, Seiberg
and Morrison from the field theory point of view. They found that
the answer is yes if the number of flavors satisfies an
upper bound. Aharony and Hanany reproduced this answer for $SU$ groups in
the brane picture. There is a slight discrepancy for $SU(2)$, which
we will discuss later. The point is that one has to avoid configurations
like in figure 3,
which we again borrowed from \cite{ahha}.

\vspace{.5cm}
\fig{An attempt to construct the $SU(2)$ theory with $N_f=5$.
Figure borrowed from \protect \cite{ahha}}
{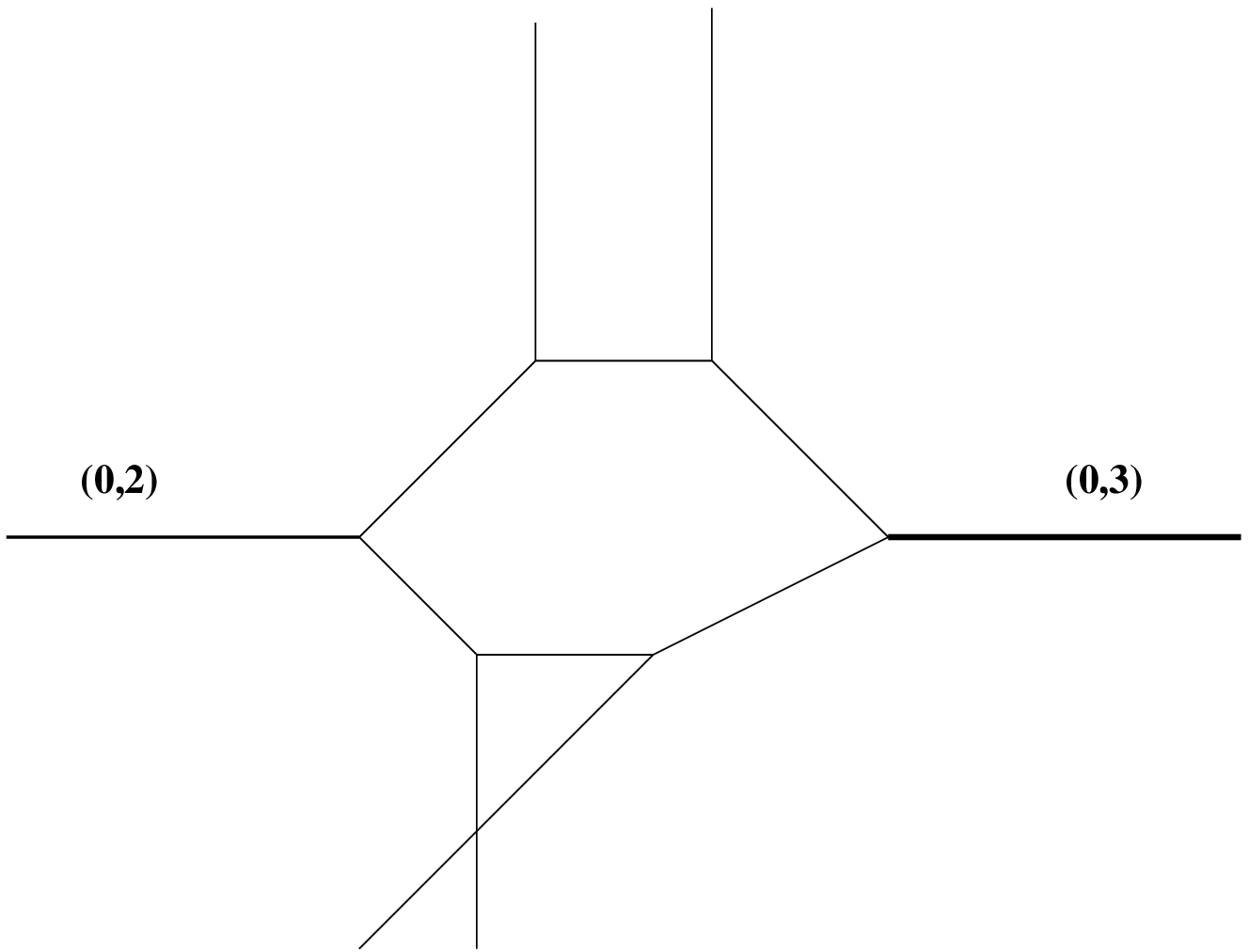}{11 truecm}
\vspace{1cm}

Even though we started out with two parallel D5 branes, the
resulting gauge theory is no longer $SU(2)$ since we get additional degrees
of freedom from the finite branes in the triangle in the lower part
of the picture. To avoid configurations like this, the branes extending
to infinity in the positive or negative 5 direction should tend away
from each other to avoid additional intersection points like
in the example discussed above. The limiting case is if the branes
come in parallel from positive infinity and leave parallel to negative
infinity. Any additional flavor (a semi-infinite brane stretching purely
in the 6 direction) would bend them towards each other and
enforce an additional intersection point. 
It is easy to see, that
this limit of parallel branes corresponds to having $N_c$ flavors
from the left and $N_c$ flavors from the right.
Taking the coupling to zero (i.e take the parallel
branes to be on top of each other) in these configurations we get
an interacting fixed point.
Hence we
get the condition $N_f \leq 2N_c$ for the existence of a brane configuration
corresponding to a non-trivial fixed point for a given $SU(N_c)$ gauge
group. This corresponds exactly to the condition found in \cite{inmosei}
for $N_c \geq 3$. The $N_c=2$ case is exceptional and allows
also fixed points for $N_f= 5,6,7,8$. They have to correspond
to configurations that do not correspond to an $SU(2)$ gauge theory
in the obvious manner described so far. Note that it would
be worse if we would have found a fixed point that corresponded to
a gauge theory which is not allowed to have a fixed
point from gauge theory considerations. While the problem
encountered in the $N_c=2$ case is just a matter of completeness (not
all gauge theory fixed points can be realized as a brane configuration
in an obvious way) the latter is a matter of consistency. It would
be in contradiction to the
assumption that all configurations of this type lead to non-trivial fixed
points.

\subsection{$SO$ and $Sp$ groups}

As a consistency check of the construction of \cite{ahha} we have
reviewed so far we will generalize the setup to $SO$ and $Sp$ groups.
We again want to analyze the conditions on the matter content
of a given gauge group that has to be satisfied in order to
have a brane construction that correponds to this gauge theory
in the obvious way described above and develops a non-trivial fixed point
at infinite coupling. We again find an upper bound on the
number of flavors which correponds exactly to the one
found in the field theory analysis of \cite{inmosei}.

As discussed by various authors \cite{ooguri,zwieb,shapere,EKG2} $Sp$ and $SO$ groups can be
realized by including an orientifold plane along the suspended color
d branes, hence in our case an O5 plane. The charge of the O5
is $\pm 1$ the charge of a physical D brane, depending on the sign of
$\Omega^2$ \cite{polchinski}. For $\Omega^2=1$ the gauge group associated to $N_c/2$
physical D branes on top of the O5 is $SO(N_c)$ and the charge of the
O5 is $-1$. For $\Omega^2=-1$ we get a $Sp(N_c)$ \footnote{To be 
able to treat $SO$ and $Sp$ groups on an equal footing and
not to introduce an additional factor of 2 we use the convention, that
$Sp(N_c)$ denotes the group with the $N_c$ dimensional fundamental
representation, $SU(2)$ is $Sp(2)$. Hence the $Sp$
groups make only sense for even $N_c$ while the $SO$ groups
are also defined for odd $N_c$. The reader should keep this in mind, since
we won't mention it anymore from now on.}
gauge theory and the O5 charge is $+1$. Whenever the O5 crosses an NS
brane $\Omega^2$ changes sign and hence also the charge \cite{shapere}. This
way the orientifold contributes to the bending of branes. Without
this sign flip it would just pull equally from both sides and leave the
orientation of the NS brane unchanged. But taking
into account the sign flip, it pushes from one side and pulls from the
other side, effectively increasing the number of physical branes by one on the
one side and decreasing it by one on the other side.

Consider the case where $\Omega^2=+1 (-1)$ between the NS branes \footnote{We 
have to choose our configuration in such a way, that at the point
where the O5 crosses the part of the 5 brane polymer extended in the 5
direction, the 5 brane is actually (1,0), i.e a NS brane. Otherwise
the polymer would not have a mirror symmetry with
the O5 as a symmetry axis.}
For $N_c/2$ color D5 branes we get a $SO(N_c)$ ($Sp(N_c)$)
gauge theory from strings stretching between the branes and their
mirrors. Each flavor brane (semi-infinite 5 brane to the left or right)
gives us the equivalence of 2 chiral multiplets in $d=4$, $N=1$, that
is one hypermultiplet. 
The charge of the O5 brane is the opposite (the same) as the
charge of a physical brane, that is a brane-mirror pair
between the NS branes, in the interior, and the same (the opposite)
left of the left and right of the right NS brane, in the exterior.
The limiting case discussed in the last section (parallel branes
extending to positive and negative infinity in the 5 direction) is
achieved, if for every NS brane the total interior
charge (D5 branes and O5) is the same as the total
exterior charge. If $n_l$ denotes the number of flavor branes from
the left and $n_r$ the number of flavor branes from the right,
this tells us:
$$n_l+1=n_r+1=N_c/2-1$$
$$(n_l-1=n_r-1=N_c/2+1)$$
From this we obtain the condition that $N_f=n_l+n_r$, the total
number of hypermultiplets, has to satisfy the relation $N_f \leq N_c-4$
($N_f \leq N_c+4$) for $SO(N_c)$ ($Sp(N_c)$) gauge group in order to
get a corresponding brane construction of a non-trivial fixed point.
These coincide precisely with the conditions found from pure
field theoretic arguments in \cite{inmosei}.

\section{The 6d analysis}
\subsection{The building blocks}

There is still one more T-duality we can apply without
destroying our NS 5 branes, namely T-dualizing the 5 direction. This
takes us once more to type IIA and the following setup of branes:
two NS 5 branes as usual along the 012345 coordinates. Stretched between
them $N_c$ D6 branes with a worldvolume along the 0123456 coordinates.
In addition there are $n_l$ semi-infinite 6 branes ending
on the left NS brane and $n_r$ semi-infinite 6 branes ending
on the right. We will again call them the $N_f=n_l+n_r$ flavor
branes. The configuration is shown schematically in figure 4.

\vspace{1cm}
\fig{The brane configuration under consideration, giving rise
to a 6 dimensional field theory. Horizontal lines represent
D6 branes, the crosses represent NS 5 branes.}{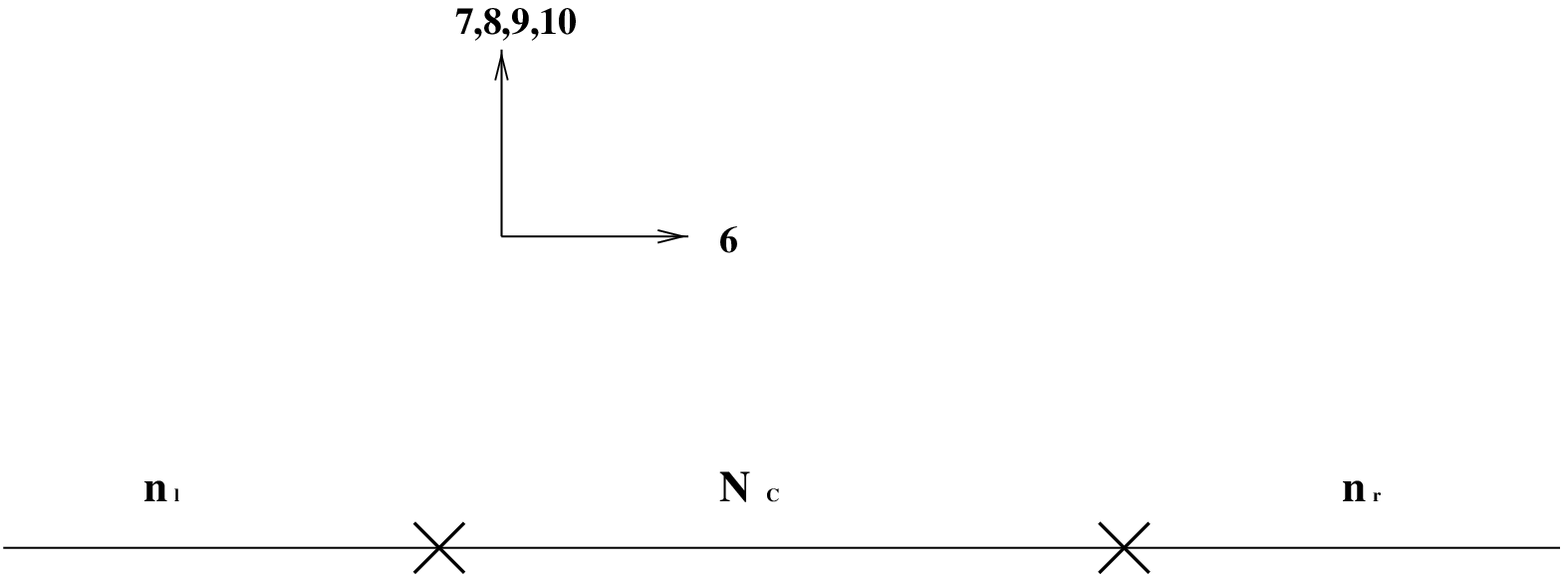}{13 truecm}
\vspace{1cm}

As discussed earlier, there is no way the 5 branes can
support a nonvanishing charge from the ends of the 6 branes,
since the charged objects, the ends of the 6 branes, are again
5 branes filling the whole worldvolume. Therefore we must
have $n_l=n_r=N_c$ and hence $N_f=2N_c$. This condition
should arise in the low energy field theory as an
anomaly requirement. This is precisely what happens, as we will
see later.
We can do the same analysis for $SO$ ($Sp$) groups. Since
the case we are interested in is again the case, where the
total charge on the left of an given NS brane is the same
as on the right, we can almost copy the calculation from
the end of the last section. The only difference is that
we are now dealing with O6 orientifolds instead of
O5 orientifolds. Taking into account that the charge of
the O6 is twice that of the O5, we obtain the conditions
$N_f = N_c -8$ ($N_f=N_c+8$) for $SO(N_c)$ ($Sp(N_c)$) gauge
group respectively.

\subsection{The low energy field theory}

To get the low energy field theory corresponding to this brane
setup we should again look for the lowest dimensional objects.
These are as usual the finite color branes, but this time we get
an equally important contribution from the full NS branes.
Since the configuration is a T-dual of the Hanany-Witten setup,
we know that we are dealing with a $N=1$ supersymmetric gauge theory
in six dimensions. Let us analyze the contributions from
the various branes. The IIA NS brane can be viewed as a
M-theory 5 brane. We know that the field theory living
on its worldvolume is that of a chiral $N=2$
supersymmetric tensor multiplet.
By introducing
the 6 branes we break $N=2$ to $N=1$. The $N=2$ tensor multiplet
decomposes into an $N=1$ tensor and a hypermultiplet.
Scalar fields correspond to fluctuations of the branes.
The $N=2$ tensor has 5 scalars which are associated to motions
of the brane in the 6,7,8,9,10 directions. Four of these
scalars are part of the $N=1$ hyper, the fifth scalar is
part of the $N=1$ tensor.
In our setup the worldvolume of the NS brane coincides with the
infinite directions of the worldvolume of the sixbrane, so there
is no room for fluctuations and the hypermultiplet does not
contribute to the massless spectrum of the theory.
There is a tensor living on both of the NS branes. Effectively,
one  tensor decouples from the theory because the scalar
in one tensor multiplet parametrizes the center of mass
motion of the two fivebranes. The scalar in the remaining
tensor is related to the distance between the two NS branes.
If the two NS branes come together, we arrive at a point where
a non-trivial fixed point is possible. The tension of strings
arising from membranes ending on the 5 branes is proportional
to the expectation value of the scalar in the remaining tensor multiplet
and thus vanishes at the fixed point.
So we are looking at a theory with one tensor multiplet, vector
and hypermultiplets. In the previous section we have seen that the brane 
setup allows only a limited amount of matter. 
From a field theory point of view these restrictions can be
reproduced by an analysis of the anomaly \cite{sweden}.
The anomaly arising from vector and hypermultiplets is
\begin{equation} \label{anomaly}
I=\alpha \tr F^4 + c (\tr F^2)^2
\end{equation}
where $\tr$ is the trace in the fundamental representation.
In the case $\alpha = 0, c>0$ the anomaly can be cancelled by
introducing a tensor multiplet. So these are the field theories
which we can compare with the theory arising from our brane configuration.
If the gauge group of the field theory has an independent fourth
order Casimir element (as is the case for $SU(N)$ with $N\geq 4$ and
$SO(N)$ with $N \geq 5$ and $Sp(N)$ with $N \geq 4$) the two conditions
$\alpha = 0$ and $ c >0$ have to be solved independently. The result
is \cite{sweden} that for the $SU$ groups only the matter content
$N_f = 2N_c$ leads to a consistent theory. For $SO$ the only possibility
is $N_f = N_c -8$ and for $Sp$ we can only have $ N_f = N+8 $. This is
in agreement with the brane analysis. However, if the gauge group
does not have an independent fourth order Casimir element ($SU(2),\, SU(3)$)
we only have to satisfy the condition $c>0$. The anomaly analysis
gives only an upper bound on the matter content.
Considering also the global anomaly \cite{vafa} for $SU(2)$ only the cases
$N_f=4,10,16$ and for $SU(3)$ only the cases $N_f=0,6,12$ are allowed.
We cannot see the
additional possibilities in our brane picture. This is however not
surprising because also in 5 dimensions the results obtained from
the brane picture and the field theory analysis did not agree in the
case of $SU(2)$.

We can also find the distance between the branes in the field theory
picture. We have already mentioned that it is related to the expectation
value of the scalar in the tensor multiplet. The interactions of this
scalar to the gauge field is described by the following terms in a
Lagrangian:
$$
\frac{1}{g^2} F^2_{\mu \nu} + (\partial \Phi)^2 + \sqrt{c} \Phi F_{\mu \nu}^2
$$
As remarked in \cite{seiberg} it is possible to absorb the coupling
$\frac{1}{g^2}$ in $\Phi$. The bosonic terms are
$$
(\partial \Phi)^2 + \sqrt{c} \Phi F_{\mu \nu}^2
$$
We obtain an effective coupling 
$$
\frac{1}{g^2_{eff}} = \sqrt{c} \Phi
$$
It is this product which we see in the brane picture. Note that this is
different in 5 dimensions, where one can also include a bare coupling:

$$\frac{1}{g^2_{eff}}=\frac{1}{g_0^2}+ c \Phi .$$

In the brane picture the bare coupling can be identified with
the distance between the NS 5 branes at the point in moduli space, where
the D 5 branes coincide. The constant $c$ determines the tension
of the magnetic BPS strings and hence is associated with the
area of the rectangle between the branes at infinite bare coupling\footnote{The BPS strings arise
from 3 branes spanning the rectangle}.

\section{Conclusion}

We showed that by applying T-duality to the Hanany-Witten setup one can
get a brane description of field theories in up to 6 dimensions. We
constructed a brane realization of various six dimensional field theories
yielding non-trivial infrared fixed points. 
Various results already known from field theory can be reproduced in 
the brane language.
Consistency of the
brane configuration gets mapped to anomaly cancellation in the
field theory analysis.

\section*{Acknowledgments}
We would like to thank B. Andreas, G. Curio, D. L\"ust and especially
K. Behrndt and T. Mohaupt for useful discussions. The work of
both of us is supported by the DFG.

\end{document}